# Quantum coherent control of a hybrid superconducting circuit made with graphene-based van der Waals heterostructures

Joel I-Jan Wang[1,†,*], Daniel Rodan-Legrain[2,†], Landry Bretheau[3], Daniel L. Campbell[1], Bharath Kannan[1,4], David Kim[5], Morten Kjaergaard[1], Philip Krantz[1], Gabriel O. Samach[4,5], Fei Yan[1], Jonilyn L. Yoder[5], Kenji Watanabe[6], Takashi Taniguchi[6], Terry P. Orlando[1,4], Simon Gustavsson[1], Pablo Jarillo-Herrero[2,*], William D. Oliver[1,2,5,*]

**Affiliations:**
[1]Research Laboratory of Electronics, Massachusetts Institute of Technology, Cambridge, MA 02139, USA.
[2]Department of Physics, Massachusetts Institute of Technology, Cambridge, MA 02139, USA.
[3]Laboratoire des Solides Irradiés, Ecole Polytechnique, CNRS, CEA, 91128 Palaiseau, France.
[4]Department of Electrical Engineering and Computer Science, Massachusetts Institute of Technology, Cambridge, MA 02139, USA.
[5]Massachusetts Institute of Technology (MIT) Lincoln Laboratory, 244 Wood Street, Lexington, MA 02421, USA.
[6]Advanced Materials Laboratory, National Institute for Materials Science, 1-1 Namiki, Tsukuba 305-0044, Japan.

[†]These authors contribute equally to this work.
[*]Correspondence to: joelwang@mit.edu; pjarillo@mit.edu; william.oliver@mit.edu

## Abstract

Quantum coherence and control is foundational to the science and engineering of quantum systems[1,2]. In van der Waals (vdW) materials, the collective coherent behavior of carriers has been probed successfully by transport measurements[3–6]. However, temporal coherence and control, as exemplified by manipulating a single quantum degree of freedom, remains to be verified. Here we demonstrate such coherence and control of a superconducting circuit incorporating graphene-based Josephson junctions. Furthermore, we show that this device can be operated as a voltage-tunable transmon qubit[7–9], whose spectrum reflects the electronic properties of massless Dirac fermions traveling ballistically[4,5]. In addition to the potential for advancing extensible quantum computing technology, our results represent a new approach to studying vdW materials using microwave photons in coherent quantum circuits.

**Main**

Since the first observation of quantum oscillations in a single-Cooper-pair box [1], superconducting qubit technology has evolved to become a leading modality for quantum computation [10], with a six-orders-of-magnitude-increase in coherence times resulting from the development of new circuit design concepts [11,12], improved material quality, and refined fabrication techniques (for a review, see Ref. [13]). Hybrid superconducting circuits also provide a platform for hosting other materials or devices with alternative electronic or mechanical degrees of freedom to access their coherent characteristics and properties via interactions with artificial atoms (qubits) or microwave photons [14,15]. Given this relevance of hybrid superconducting circuits to both quantum computing technologies and the fundamental sciences, it is essential to expand the scope of the field by introducing new materials and experimental approaches.

One highly anticipated development is the application of van der Waals (vdW) materials [16] – a family of layered materials including semi-metals, insulators, semiconductors, ferromagnetic materials [17], superconductors, and topological insulators – in the context of quantum information science [18–20]. These materials can be assembled in specific arrangements to create new electronic devices referred to as vdW heterostructures [16]. While vdW heterostructures have been studied intensively using DC transport measurements, to date, only a few very recent experimental studies on vdW materials integrated into circuit quantum electrodynamics (cQED) systems have been reported [21,22]. More importantly, the literature does not include any accounts of temporal quantum coherence or quantum control with these material systems, and both are prerequisites for quantum information science and technology applications.

Figure 1a shows the schematic of a hybrid superconducting circuit made with graphene-based vdW heterostructures. Graphene, an atomic layer of carbon atoms in a hexagonal lattice [23], can inherit superconductivity through the proximity effect [3,6]. When encapsulated in hexagonal boron nitride (hBN), the superconductor-graphene-superconductor (S-G-S) junction supports a voltage-tunable, bipolar Josephson current in the ballistic regime [4,5,24]. In our experiment, the S-G-S junctions are built into a voltage-controlled transmon circuit that enables us to perform both frequency-resolved spectroscopy and coherent quantum control using cQED techniques (Fig.1a).

The voltage-controlled transmon, also known as a gatemon, employs superconductor-normal conductor-superconductor junctions as the non-linear element in a superconducting circuit. Voltage-controlled transmons have been realized with a semiconductor nanowire[7,8] and a 2-dimensional electron gas [9] proximitized by epitaxially grown superconductors, and, over time, have achieved temporal coherence comparable to that of typical transmon qubits in the cQED architecture based on superconductor – insulator – superconductor junctions [8,25]. While the graphene-based gatemon shares the same voltage tunability as its semiconductor-based counterparts, it may as well offer complementary benefits to the existing scheme provided by its unique electronic properties and the extensibility to complex heterostructures in high-quality-factor (low dissipation) superconducting circuits. In addition to its relevance for extensible quantum computing [7,26], this particular qubit configuration enables simultaneous access to time- and energy-domain information pertaining to the underlying quantum conductor, the electronic properties of which can be largely tuned by the gate voltage.

To make the hybrid device, we apply the polymer-based dry pick-up and transfer techniques (Fig. 1b) [6,27] to assemble the vdW heterostructures, which, from bottom to top, consist of a thick hBN substrate, a monolayer graphene sheet, and a top hBN layer. The bottom hBN layer serves as an atomically flat, ultra-pristine gate dielectric [16], which allows access to a wide carrier density range through electrostatic gating. The top hBN layer encapsulates the graphene, thereby protecting it from becoming contaminated in subsequent steps. The assembled vdW stack is transferred onto an aluminium backgate on the qubit substrate, a high-resistivity silicon chip coated with high-quality aluminium and patterned into a coplanar wave guide, shunting capacitors, DC bias lines, and the ground plane (Fig. 1d). The aluminium superconducting electrodes of the S-G-S junctions couple to the encapsulated graphene flakes via 1-D edge channels [27], and make transparent connections to the ground plane and capacitors (details about fabrication and measurement setup can be found in **Methods**).

Without loss of generality, we center our discussion on the device shown in Fig. 1c (Qubit 1 in Fig. 1a), on which all measurements discussed in this paper were taken, while the backgate of the other qubit on the same chip was fixed at 5V (measurements from additional devices can be found in supplementary information figure 1, 2 and 3). The charging energy $E_C$ of this device is $E_C/h$ = 100 MHz, as determined based on an electromagnetic simulation of the shunting capacitor.

First, we measure how the transmission of the resonator responds to a varying gate voltage $V_g$ by scanning the readout frequency $f_{ro}$ around the bare-resonator frequency $f_{R\_bare}$ = 7.34 GHz. In figure 2a, the resonator spectrum is plotted as a function of $f_{ro}$ and $V_g$. We observe a splitting of the cavity resonance frequency $f_R$ at $V_g \sim$ -2.2V and $V_g \sim$ -2.8V, indicating hybridization between the resonator and the qubit in the strong coupling regime.

The cQED approach enables qubit readout by monitoring the qubit-state-dependent transmission through the resonator. In addition to the readout tone, another driving pulse $f_{dr}$ is applied to induce state transitions in the qubit. Figure 2b (main panel) shows such a qubit spectrum plotted against the driving frequency $f_{dr}$ and $V_g$. The distinct peak position indicates the qubit frequency $f_{qb}$, which corresponds to the transition energy $E_{01} = h*f_{qb}$ at any given gate voltage. We now discuss major features of this spectrum.

The qubit frequency $f_{qb}$ exhibits a large tunability range from 6 GHz to 12 GHz. In transmon qubits, the qubit frequency is generally given by $f_{qb} = E_{01}/h \sim \sqrt{8E_C E_J(V_g)}/h - E_C/h$ [11], where $E_{01}$ is the energy corresponding to the transition from ground state to the first excited state, $E_J$ the Josephson coupling energy of the junction [28], and $h$ the Plank's constant. With $E_C/h$ = 100 MHz used for this circuit, the critical Josephson current $I_c$ of the S-G-S junction is estimated to be 90 nA ~ 360 nA, consistent with results reported in previous DC-transport studies [3]. In contrast to superconductor – insulator – superconductor tunnel junctions, where $E_J$ is generally modified using a current to induce a magnetic field in a SQUID-type configuration, in superconductor-normal conductor-superconductor devices the value of $E_J$ can be tuned with $V_g$ due to two effects: (i) $V_g$ changes the Fermi energy and thereby directly modifies the density of states, and therefore the total number of transmission channels [6,23,29], and (ii) $V_g$ modifies the transmission probability for a number of high-transmission conduction channels [7,8,30]. Overall this results in an electrostatic control of the Andreev spectrum in graphene [6] and the corresponding Josephson coupling energy $E_J$ [28]. The qubit spectrum has a minimum at $V_g$ = -2.52 V, where the Fermi level reaches the point with minimum density of states (minimum $E_J$) and is identified as the charge neutrality point ($V_{CNP}$) of the graphene qubit. As a gapless semi-metal, graphene supports the Josephson effect in the N-doped ($V_g > V_{CNP}$) region with electron-like charge carriers, and in the P-doped ($V_g < V_{CNP}$) region with hole-like carriers. The qubit spectrum shows clear asymmetry in $f_{qb}$ with respect to $V_{CNP,}$ manifested as a lower $f_{qb}$ value in the P-doped region for a given gate voltage – as compared to its equal carrier density (same $|V_g - V_{CNP}|$) counterpart in the N-doped region – and pronounced $f_{qb}$ oscillations observed in the P-doped region. This asymmetry arises from the N-type doping provided by the Ti/Al electrodes [31], as evidenced by the negative $V_{CNP}$ value of our device. In the vicinity of the electrode, doping from electrodes outweighs the effect of negative $V_g$ and yields an N-P-N potential profile within the graphene (Fig. 2b, bottom panel), resulting in two semi-transparent interfaces (P-N junctions) for charge carriers. In addition to suppressing the magnitude of $f_{qb}$ [29], the doping – and the resulting interfaces – create an electronic equivalence of a Fabry–Pérot cavity with an effective cavity length $L_C$ inside the S-G-S junction. Furthermore, as a function of $V_g$, the Fermi wavevector $k_F \propto \sqrt{V_g}$ of the carriers inside this cavity vary. This variation is thus manifested in the local maxima (minima) of $f_{qb}$, where the resonant transmission (reflection) condition for adjacent peaks $\Delta k_F \cdot L_C = n\pi$ for integer $n$ is satisfied. From the peak-to-peak separation in the gate voltage ($\Delta V_g$ in Fig. 2b), we extract the $L_C$ for our device as ~ 110 nm, accounting ~ 200 nm in total for the regions doped by the metallic electrodes [4,5]. In addition to observing the Fabry–Pérot oscillations, we note an aperiodic fluctuation of $f_{qb}$ with an amplitude of ~ 100 MHz in the P-doped region. This reproducible (over weeks of measurement) fluctuation might be associated with universal conductance fluctuation in mesoscopic Josephson weak links [32].

Next, we demonstrate that the device can be operated as an artificial atom by performing qubit operations in the time domain. Fig. 3a shows measurements of the qubit state while varying the drive frequency and the pulse duration $\tau_{Rabi}$ for four different powers, measured at a fixed Vg = -4.38 V. When $f_{dr} = f_{qb}$ (8.811 GHz), the pulse induces Rabi oscillations of the qubit. In the Bloch sphere representation, this oscillation corresponds to rotations about the x-axis and the qubit alternates between |0> and |1> states continuously. The Rabi oscillations are also used to calibrate the $R_X^{\pi}$ and $R_X^{\pi/2}$ pulses, which rotate the Bloch vector about the x-axis by $\pi$ and $\pi/2$, respectively. We apply a $R_X^{\pi}$ pulse followed by a delayed readout pulse, yielding an exponential decay trace with an energy relaxation time of $T_1 \sim 36$ ns (Fig. 3b). Fig. 3c plots the $T_1$ measurement repeated at different bias points with values that range from 12 to 36 ns.

We studied the dephasing of our graphene qubit using Ramsey interferometry techniques. In the rotating frame, the application of two slightly detuned $R_X^{\pi/2}$ pulses separated by a delay time $\tau_{Ramsey}$ flips the qubit to the x-y plane and allows the qubit to rotate about the z-axis by $\Delta f \cdot \tau_{Ramsey}$, where $\Delta f = f_{dr} - f_{qb}$ is the detuning. Figure 4a (main panel) shows Ramsey oscillations as a function of detuning and time delay at a given gate voltage. Fringes at finite detuning indicate coherent precession around the z-axis of the qubit. Fig. 4b plots the discrete-time Fourier transform of the Ramsey fringes (calculated using the fast Fourier transform (FFT) algorithm) and reveals at least three frequency components in addition to the main qubit frequency. The additional frequency components may be associated with coupling to spurious two-level systems embedded in the heterostructures [8,29] or unintentional population of higher-excited qubit states due to the relatively low qubit anharmoncity. Finally, fitting the decay envelope of the pattern at finite detuning yields $T_2^* \sim 55$ ns (Fig. 4a, middle panel). This value is close to twice $T_1$ at the same gate voltage, suggesting that our qubit coherence is currently limited by energy relaxation.

It has been shown that the supercurrent transport characteristics in proximitized graphene vary with the Fermi energy [5,6,24,33,34]. In this work, however, we do not observe a strong $V_g$-dependence of the coherence (Fig. 3c) to the extent recently observed in nanowire gatemons [8]. This strongly indicates the coherence in our device is not limited by the graphene junction. Rather, we attribute the lack of a strong $V_g$-dependence to the relatively short relaxation times $T_1$ in the present measurement, due to other decay channels in parallel with the intrinsic loss of the graphene junction. For example, it is clear that one major contribution to the relaxation in the current generation of our devices is photon loss due to geometric capacitive coupling between the qubit shunting capacitor and the backgate. This will be improved straightforwardly in future devices by redesigning the capacitor geometry and introducing on-chip low-pass filters [35].

In summary, we have successfully demonstrated temporal quantum coherence and control of a superconducting circuit based on vdW heterostructures. Our demonstration used a transmon qubit comprising S-G-S junctions. The spectrum of the qubit exhibits the characteristic electronic properties of the underlying ballistic S-G-S junction, such as voltage tunability over more than 6 GHz, and Fabry–Pérot oscillations in the P-doped region. Our results show that there is an opportunity to engineer a qubit spectrum in order to optimize charge stability and coherence. For example, a device can be designed with multiple voltage sweet spots given by the Fabry–Pérot oscillations where the dephasing is minimized [8]. With the assistance of local electrostatic gating or a superlattice band structure given by vdW heterostructures [16], it is possible to further engineer the spectrum in order to achieve different functionality. The existing rich context of graphene physics, such as the topologically protected Josephson effect [33,34,36] as well as the gate-dependent current-phase relation [24], also provide the opportunity to sense and investigate the effects of the microscopic transport processes via a single quantum degree of freedom, i.e. the temporal coherence of a qubit, supporting both fundamental discovery and new qubit design concepts. These results suggest that the graphene-based superconducting qubit has great potential for extensible superconducting quantum computing and could be a key element of topological quantum computing architectures [22,26].

Finally, we emphasize that the experiments presented in this study can be adapted for other van der Waals materials and for more complex heterostructures, the planar geometry of which is highly integrable to the cQED architecture. The extraordinary and versatile electronic properties of these heterostructures, in combination with their epitaxial precision, make vdW-based devices a promising alternative for constructing key elements of novel solid-state quantum computing platforms. A hybrid superconducting circuit with vdW materials also provides a fertile arena for studying light-matter interaction in such materials, as well as probing the electronic degrees of freedom using a non-invasive approach with high precision. Of particular interest are systems showing intrinsic superconductivity with a topological phase [37] or many-body phases [38], which can be readily incorporated into a superconducting quantum circuit similar to the one presented herein.

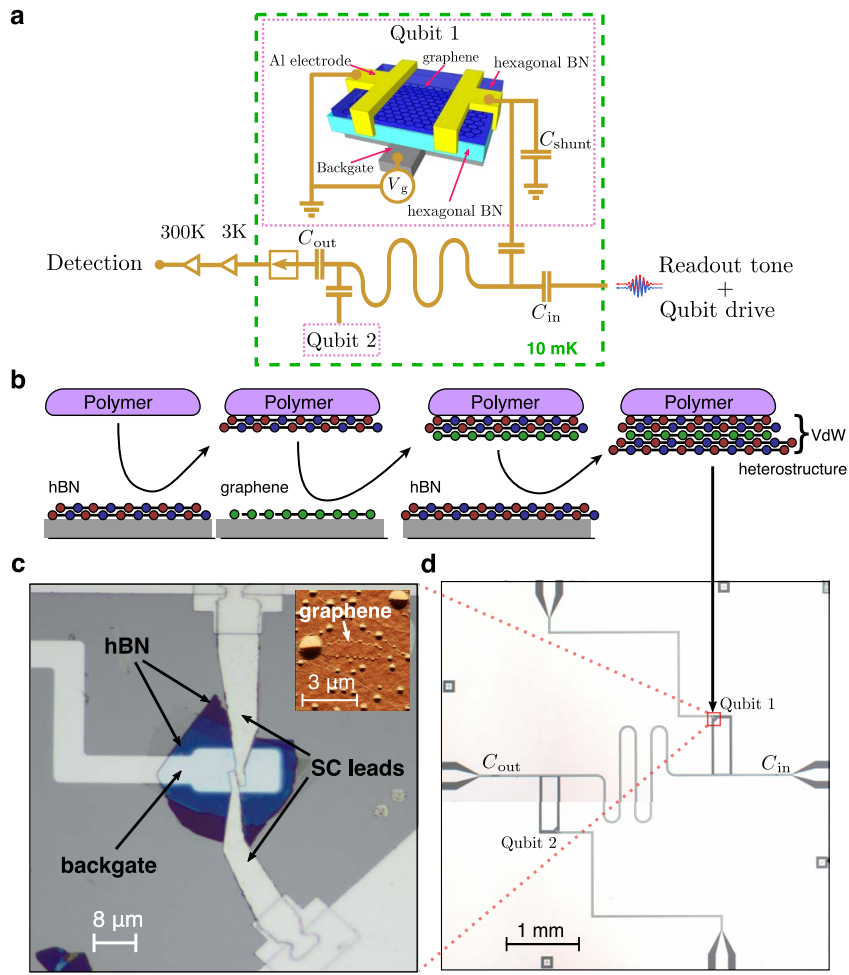

**Fig. 1. Fabrication of graphene transmon qubits**. (**a**) Schematic of the hexagonal boron nitride (hBN)-encapsulated superconductor-graphene-superconductor (S-G-S) junctions embedded in a circuit quantum electrodynamics (cQED) system. (**b**) Assembly of van der Waals heterostructures using a dry polymer-based pick-up and transfer technique. (**c**) Optical micrograph of the graphene transmon qubit. Inset: AFM micrograph of the encapsulated graphene before making electrical contact to the superconducting electrodes. (**d**) Qubit chip made of high-quality aluminium. Each shunting capacitor is cut out at the corner (red box, Fig. 1c) to host the assembled van der Waals (vdW) stack. Bonding pads on the top and bottom of the chip are used for backgate control.

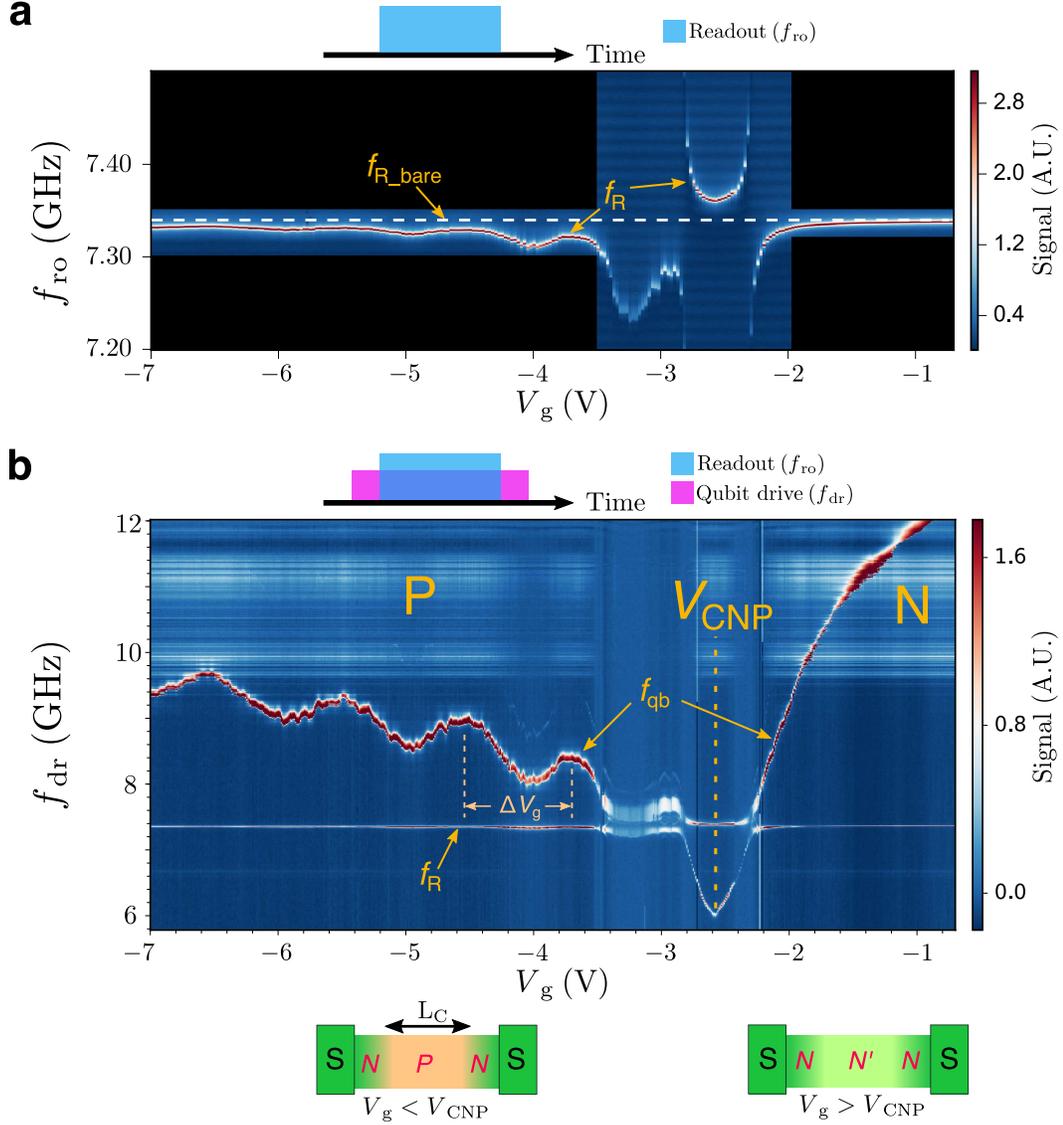

**Fig. 2. Spectroscopy of a graphene transmon qubit.** (**a**) Resonator spectrum as a function of readout frequency ($f_{ro}$) and gate voltage ($V_g$). Resonance frequency $f_R$ exhibits a splitting at $V_g \sim -2.2$V and $V_g \sim -2.8$V. The bare resonance frequency $f_{R\_bare}$ of the resonator is marked at 7.34 GHz by the white dashed line. (**b**) Qubit spectrum as a function of drive frequency $f_{dr}$ and $V_g$. Qubit frequency $f_{qb}$ is asymmetric in respect to the charge neutrality point ($V_{CNP} \sim -2.5$V). The Fabry–Pérot oscillation and an aperiodic fluctuation of $f_{qb}$ appear in the P-doped region. Bottom panel: schematics of the graphene potential profiles in P (left) and N (right) regions. Cavity length $L_C$ extracted from $\Delta V_g$ is ~110 nm. Note that the backgate of the other qubit on the same chip is fixed at 5V throughout the measurements presented in this paper.

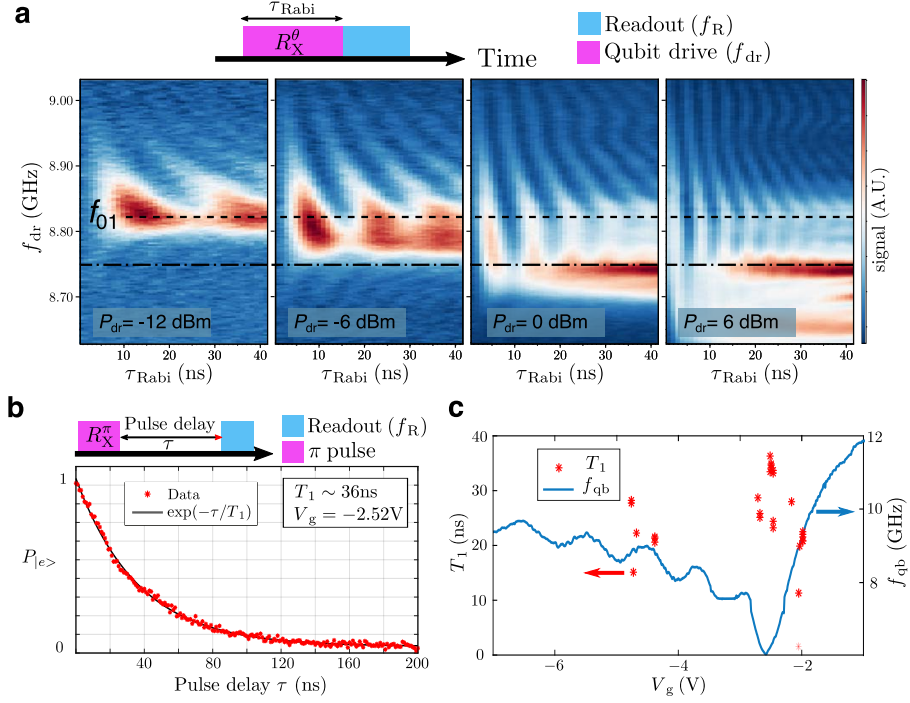

**Fig. 3. Rabi oscillation and energy relaxation of a graphene qubit.** (**a**) Rabi oscillation, associated with $f_{01}$ qubit transition, as a function of drive frequency $f_{dr}$ and pulse duration $\tau_{Rabi}$ measured with different qubit drive power values at $V_g = -4.38$V. The qubit $f_{12}$ and/or two-photon $f_{02}/2$ transition is induced at high driving power ($P_{dr} = 0$ dBm and 6 dBm), as indicated by the pronounced signal below the Rabi oscillation (lower dashed line). (**b**) Energy relaxation time $T_1$ measurement taken at $V_g = -2.52$ V using a π-pulse calibrated from the Rabi oscillation. The fitting of the qubit population to an exponential decay yields $T_1 \sim 36$ ns. (**c**) $T_1$ measured at different voltage bias points.

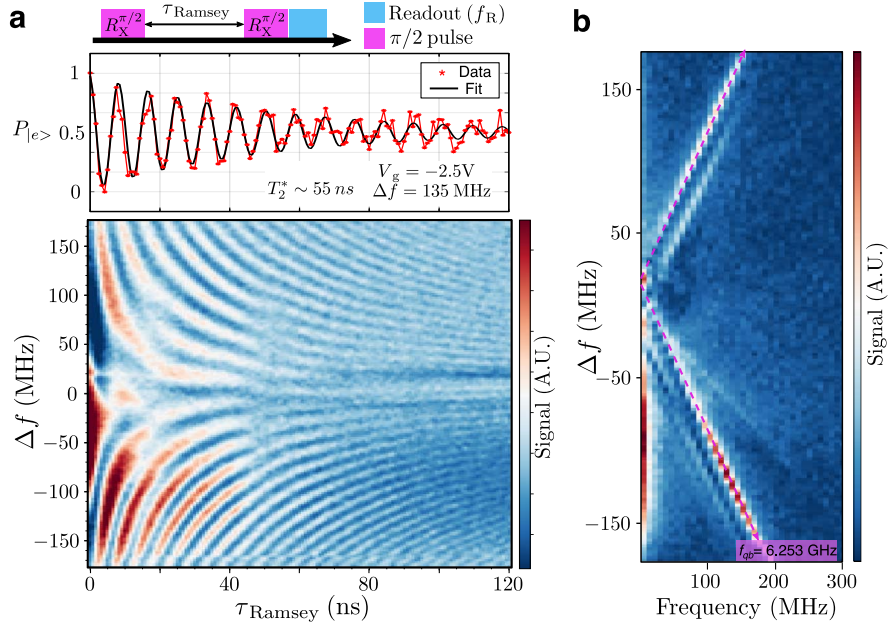

**Fig. 4. Qubit dephasing and Ramsey measurement.** (**a**) Main panel: Ramsey fringes as a function of detuning and time delay $\tau_{Ramsey}$ measured at $V_g$= -2.5 V. A repeated beating pattern appears around $|\Delta f|$< 75 MHz, separated by ~ 30 ns. Middle panel: Line cut in the Ramsey fringes with finite detuning. The dephasing time $T_2^*$~ 55 ns is obtained by fitting to the function = $\exp(-\tau_{Ramsey}/T_2^*) \cdot \cos(2\pi \cdot \Delta f \cdot \tau_{Ramsey})$ . (**b**) Discrete time Fourier transform of the Ramsey measurement. The condition $\Delta f = |f_{dr} - f_{qb}|$ is represented by the dashed purple line. Extra frequency components presented in this plot might be associated with coupling to spurious two-level system or the excitation of 2-photon transition in the heterostructures.

**Acknowledgements:**
We gratefully acknowledge M. Augeri, J. Birenbaum, P. Baldo, G. Fitch, M. Hellstrom, K. Magoon, A. Melville, P. Murphy, B. M. Niedzielski, B. Osadchy, D. Rosenberg, R. Slattery, C. Thoummaraj, and D. Volfson at MIT Lincoln Laboratory for technical assistance. This research was funded in part by the U.S. Army Research Office Grant No. W911NF-17-S-0001, and by the Assistant Secretary of Defense for Research & Engineering via MIT Lincoln Laboratory under Air Force Contract No. FA8721-05-C-0002. P.J.-H. and L.B. were partly supported by the Gordon and Betty Moore Foundation's EPiQS Initiative through Grant No. GBMF4541.This work made use of the MRSEC Shared Experimental Facilities at MIT, supported by the National Science Foundation under award number DMR-14-19807 and of Harvard CNS, supported by NSF ECCS under award no. 1541959. Growth of BN crystals was supported by the Elemental Strategy Initiative conducted by the MEXT, Japan and JSPS KAKENHI Grant Numbers JP15K21722 and JP25106006. D.R.-L. acknowledges support from Obra Social "la Caixa" Fellowship. M.K. acknowledges support from the Carlsberg Foundation. The views and conclusions contained herein are those of the authors and should not be interpreted as necessarily representing the official policies or endorsements of the US Government.

**Author Contributions**

J.I.-J.W., L.B., S.G., T.P.O., P.J.-H., and W.D.O. conceived and designed the experiment. D.R.-L. and J.I.-J.W. fabricated the graphene devices. J.I.-J.W., F.Y. and S.G. conducted the measurements and analyzed the data. D.K., G.O.S., and J.L.Y. supported sample fabrication. D.L.C., B.K., M.K., and P.K. supported measurements. K.W. and T.T. supplied the hBN crystals. J.I.-J.W., S.G. and W.D.O. wrote the manuscript. All authors discussed the results and commented on the manuscript.

**Methods**

The vdW heterostructures are made using a dry-polymer based process[6,27] followed by annealing at ~350 C in vacuum or forming gas[6,29] to remove excessive hydrocarbon compounds within the heterostructures. After transferring the stack onto the qubit chip, superconducting leads are defined by standard e-beam lithography and reactive ion etching (RIE) that facilitates the 1-D edge contact[6,27,29] to thermally evaporated titanium (7 nm) and aluminium (120 nm). A second lithography step is then applied to define areas for in-situ ion-milling and e-beam evaporation of 250 nm of aluminium that bridges the electrodes to the ground plane and shunting capacitors.

The devices are measured in a dilution refrigerator with a base temperature of approximately 10 mK. Dispersive readout and qubit control are implemented using microwave pulses applied to a half-wavelength coplanar waveguide (CPW) resonator, which capacitively couples to two qubits at opposite ends of the CPW in a standard circuit-based cavity quantum electrodynamics (cQED) setup (Figs. 1a and 1d). The backgate voltage is applied through low-pass-filtered DC-bias lines and used to tune the Fermi energy in the graphene.

Supplementary Information for

# Coherent control of a hybrid superconducting circuit made with graphene-based van der Waals heterostructures

Joel I-Jan Wang, Daniel Rodan-Legrain, Landry Bretheau, Daniel L. Campbell, Bharath Kannan, David Kim, Morten Kjaergaard, Philip Krantz, Gabriel O. Samach, Fei Yan, Jonilyn L. Yoder, Kenji Watanabe, Takashi Taniguchi, Terry P. Orlando, Simon Gustavsson, Pablo Jarillo-Herrero, William D. Oliver

**Content**

1. Spectroscopy data from additional graphene-based transmon devices (supplementary fig.1 and supplementary fig. 2).
2. Time-domain measurements from additional graphene-based transmon devices (supplementary fig.3).

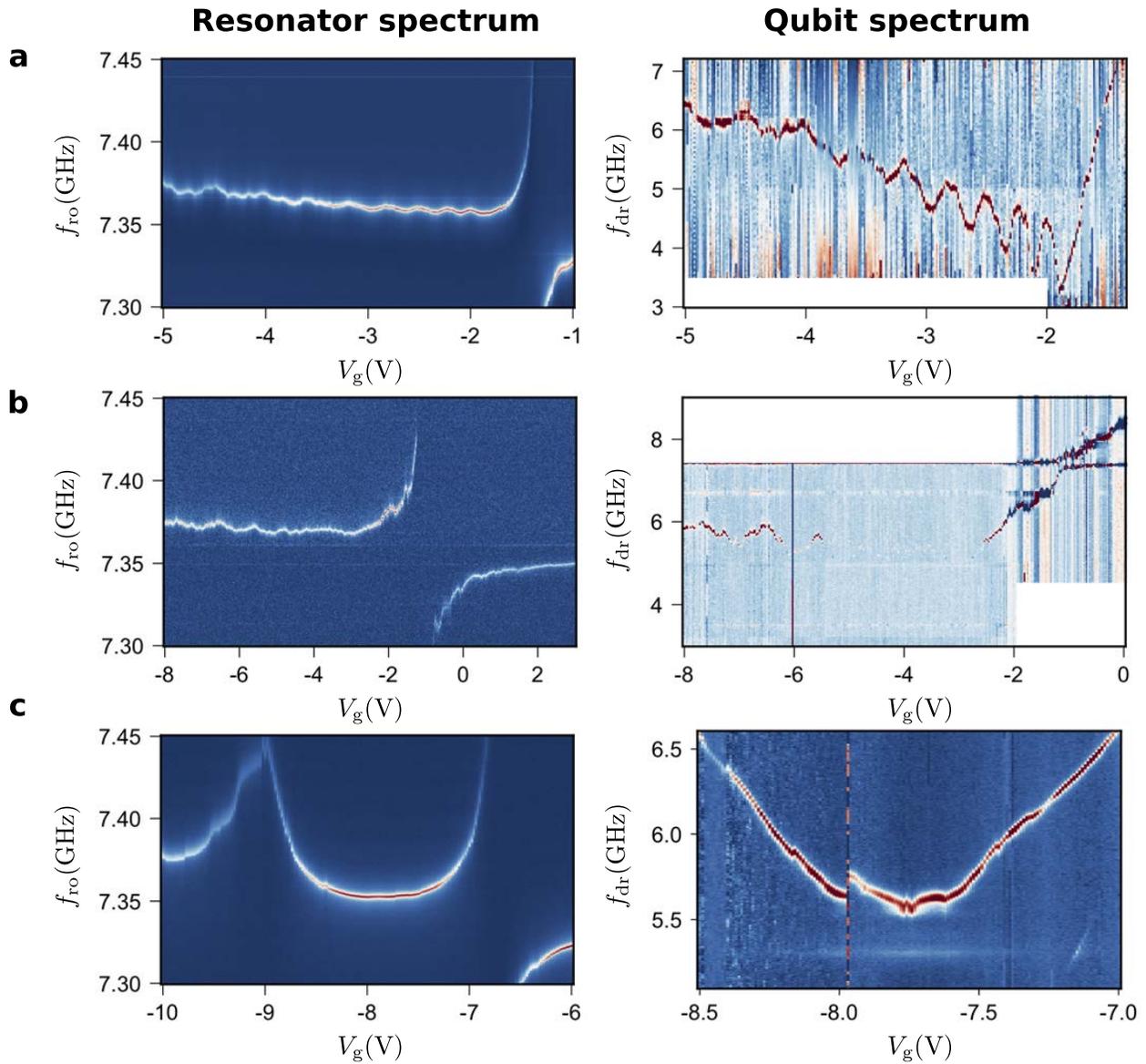

**Supplementary fig.1. Spectroscopy data from additional devices.** Resonator (left panels) and qubit spectra (right panels) as functions of backgate voltage $V_g$ from three additional devices shown in (**a**), (**b**), and (**c**) respectively. Note in device (**c**), for which the graphene is presumably highly-doped ($V_{CNP}$ ~-7.8), the Fabry–Pérot oscillation of qubit frequency is not observed. Data is plotted in arbitrary units.

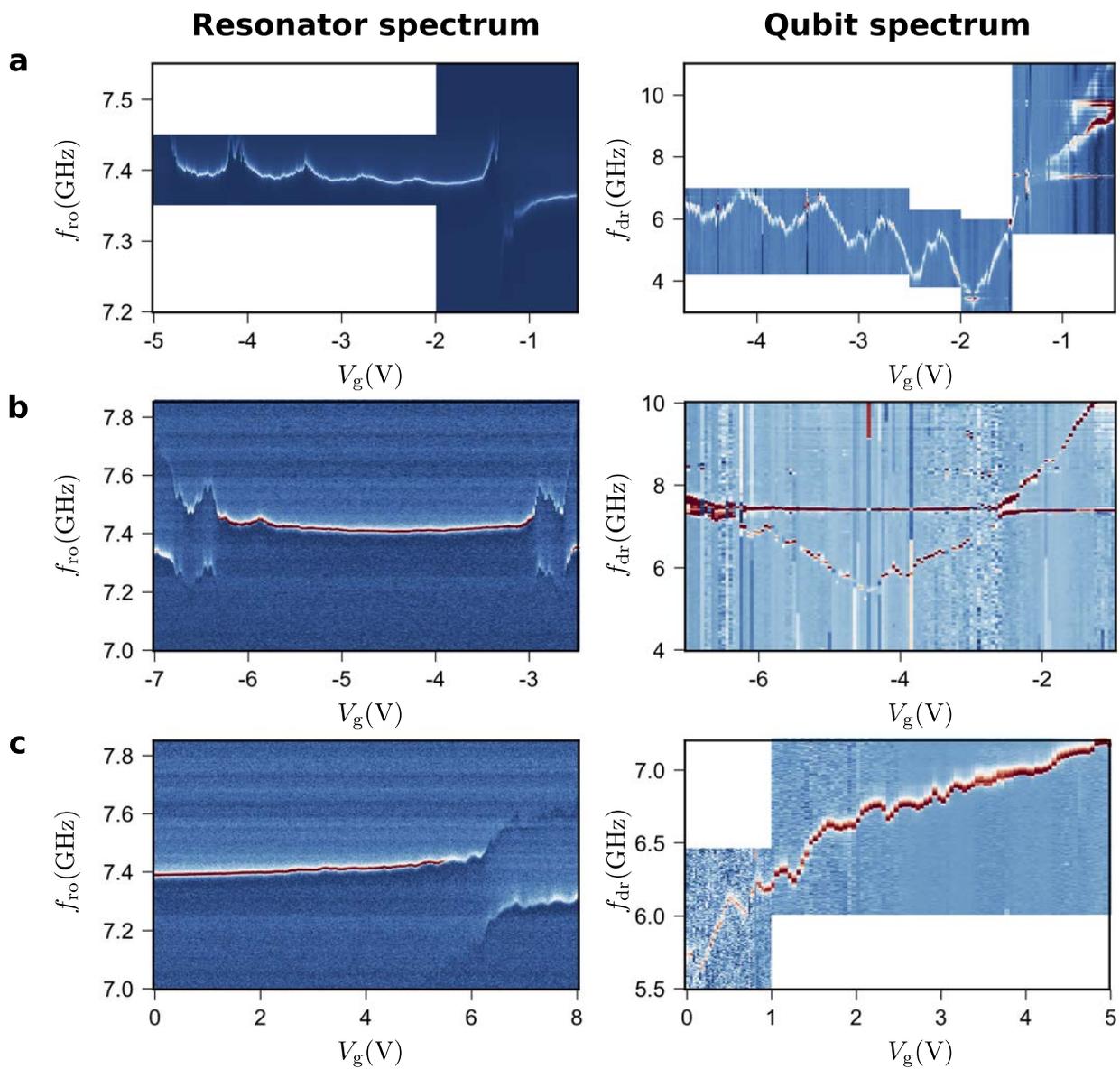

**Supplementary fig.2. Spectroscopy data from additional devices.** Resonator (left panels) and qubit spectra (right panels) as functions of backgate voltage $V_g$ from three additional devices, shown in (**a**), (**b**), and (**c**) respectively. Data is plotted in arbitrary units.

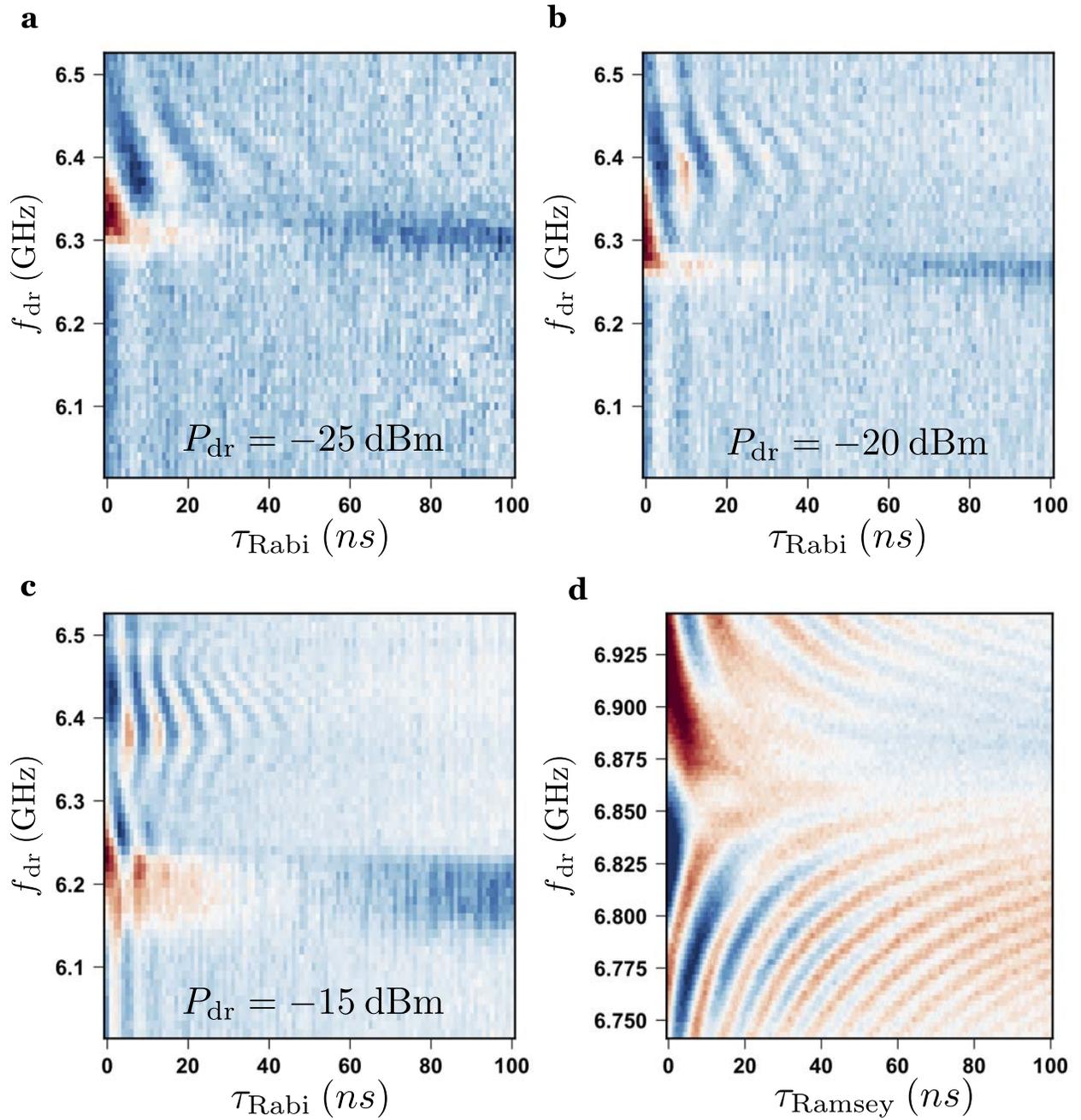

**Supplementary fig.3. Time domain measurements from additional devices [corresponding to supplementary fig.1(c)].** (**a**)-(**c**) Rabi oscillation measured with different qubit-drive power $P_{dr}$ at $V_g$= -7.59 V. (**d**) Ramsey fringes as a function of qubit-drive frequency $f_{dr}$ and time delay $\tau_{Ramsey}$ measured at $V_g$= -7.34 V. The energy relaxation time $T_1$ and dephasing time $T_2^*$ are of the same order as those reported in the main text. Data is plotted in arbitrary units.